\documentclass[12pt,reqno]{article}
\usepackage{amsfonts}
\usepackage{amsmath}
\usepackage{amsbsy}

\usepackage{amssymb,latexsym}
\usepackage{setspace}
\numberwithin{equation}{section}

\begin{document}
 \allowdisplaybreaks[1]
\title{Heterotic String Dynamics in the Solvable Lie Algebra Gauge}
\author{Nejat T. Y$\i$lmaz\\
          Department of Mathematics
and Computer Science,\\
\c{C}ankaya University,\\
\"{O}\u{g}retmenler Cad. No:14,\quad  06530,\\
 Balgat, Ankara, Turkey.\\
          \texttt{ntyilmaz@cankaya.edu.tr}}
\maketitle
\begin{abstract}A revision of the torodial Kaluza-Klein compactification of
the massless sector of the $E_{8}\times E_{8}$ heterotic string is
given. Under the solvable Lie algebra gauge the dynamics of the
$O(p,q)/(O(p)\times O(q))$ symmetric space sigma model which is
coupled to a dilaton, $N$ abelian gauge fields and the
Chern-Simons type field strength is studied in a general
formalism. The results are used to derive the bosonic matter field
equations of the massless sector of the $D$-dimensional
compactified $E_{8}\times E_{8}$ heterotic string.

\end{abstract}

\section{Introduction}
The supergravity theory which has the highest spacetime dimension
is the $D=11$, $\mathcal{N}=1$ supergravity \cite{d=11}. There are
three types of supergravity theories in ten dimensions namely the
IIA \cite{2A1,2A2,2A3}, the IIB \cite{2B1,2B2,2B3} and as the
third supergravity, the ten dimensional $\mathcal{N=}1$ type I
supergravity theory which is coupled to the Yang-Mills theory
\cite{d=10,tani15}. One can obtain the $D=10$, IIA supergravity
theory by the Kaluza-Klein dimensional reduction of the $D=11$
supergravity on the circle, $S^{1}$. The supergravity theories in
$D<10$ dimensions can be obtained from the $D=11$ and the $D=10$
supergravities by the dimensional reduction and the truncation of
fields. A general treatment of the supergravity theories can be
found in \cite{westbook,sssugradivdim,westsugra,tani}.

The ten dimensional IIA supergravity and the IIB supergravity
theories are the massless sectors or the low energy effective
limits of the type IIA and the type IIB superstring theories
respectively \cite{kiritsis}. The type I supergravity theory in
ten dimensions on the other hand is the low energy effective limit
of the type I superstring theory and the heterotic superstring
theory \cite{kiritsis}. The eleven dimensional supergravity is
conjectured to be the low energy effective limit of the eleven
dimensional M-theory.

The symmetries of the supergravity theories are important to
understand the symmetries and the duality transformations of the
string theories. Especially the global symmetries of the
supergravities give the non-perturbative U-duality symmetries of
the string theories and the M-theory \cite{nej125,nej126}. An
appropriate restriction of the global symmetry group $G$ of the
supergravity theory to the integers ${\Bbb{Z}}$, namely
$G({\Bbb{Z}})$, is conjectured to be the U-duality symmetry of the
relative string theory which unifies the T-duality and the
S-duality \cite{kiritsis,nej125}.

In supergravity theories which possess scalar fields, the global
(rigid) symmetries of the scalar sector can be extended to the
other fields as well, thus the global symmetry of the scalars will
be the global symmetry of the entire bosonic sector of the theory.
For the majority of the supergravities, the scalar manifolds are
homogeneous symmetric spaces which are in the form of cosets $G/K$
\cite{fre}, and the scalar lagrangians can be formulated as the
non-linear coset sigma models, in particular the symmetric space
sigma models. The dimension of the coset space $G/K$ is equal to
the number of the coset scalars of the theory. In general $G$ is a
real form of a non-compact semi-simple Lie group and $K$ is its
maximal compact subgroup. Thus the coset $G/K$ can locally be
parameterized through the exponential map by using the solvable
Lie subalgebra of $G$ \cite{hel}. This parametrization is an
effective tool for the dynamics of the supergravity theories and
it is called the solvable Lie algebra gauge \cite{fre}.

 When we apply
the Kaluza-Klein dimensional reduction to the bosonic sector of
the $D=11$ supergravity \cite{d=11} over the tori $T^{n}$, where
$n=11-D$, we obtain the $D$-dimensional maximal supergravity
theories \cite{julia1,julia2,pope}. The global (rigid) symmetry
groups of the bosonic sectors of the reduced lagrangians
 are in split real form (maximally non-compact). In other words for the scalar coset
 manifolds, $G/K$
of the maximal supergravities, $G$ which is the global symmetry
group, is a semi-simple split real form and
 $K$ is its maximal compact subgroup. Thus the coset spaces $G/K$
 can be  parameterized by the
Borel gauge which is a special case of the solvable Lie algebra
gauge. Therefore the scalar sectors of the maximal supergravities
can be formulated as symmetric space sigma models. However we
should remark that in certain dimensions in order to formulate the
scalar sectors as symmetric space sigma models one has to make use
of the dualisation methods to replace the higher-order fields with
the newly defined scalars. On the other hand when one considers
the Kaluza-Klein compactification of the bosonic sector of the ten
dimensional $\mathcal{N=}1$ simple supergravity that is coupled to
$N$ abelian gauge multiplets on the tori $T^{10-D}$
\cite{d=10,tani15}, one can show that when a single scalar is
decoupled from the others, the rest of the scalars of the lower,
$D$-dimensional theories can be formulated as $G/K$ symmetric
space sigma models \cite{heterotic}. One can even enlarge the
coset formulations of the scalars by using partial dualisations.
Unlike the maximal supergravities for this class of
supergravities, the global symmetry group $G$ is not necessarily a
split real form but it is in general a semi-simple, non-compact
real form and again $K$ is a maximal compact subgroup of $G$.
Therefore in this case one makes use of the more general solvable
Lie algebra gauge \cite{fre} to parameterize the scalar coset
manifolds for the reduced theories.

In this work considering the achievements of \cite{heterotic} we
focus on the dynamics of the toroidally compactified low energy
effective $E_{8}\times E_{8}$ heterotic string in the solvable Lie
algebra gauge. In accordance with the formulation of
\cite{heterotic} we will first show how one can obtain a lower
dimensional bosonic lagrangian starting from the ten dimensional
$\mathcal{N=}1$ type I supergravity theory which is coupled to the
abelian Yang-Mills theory \cite{d=10,tani15}. We will formulate
the $D$-dimensional bosonic lagrangian in a compact form in which
the terms governing the scalars and the coupling abelian gauge
fields are constructed in the solvable Lie algebra gauge. Then in
a general formalism we will work out in detail the bosonic field
equations of the symmetric space sigma model which is coupled to a
dilaton, an arbitrary number of abelian gauge fields and a
two-form field whose kinetic term is expressed in terms of its
Chern-Simons field strength. Our formulation will be a generalized
one which uses arbitrary coupling constants. Also it is performed
for a general, $O(p,q)$ global symmetry group of the scalar coset.
In our generalized formulation we will make use of the
construction of \cite{nej3} to derive the field equations of the
coset scalars. To denote the relevance of our construction with
the supergravity theories we will give three examples of
Maxwell-Einstein supergravities on which the derivation we perform
is applicable and whose bosonic field equations correspond to the
ones we derive in algebraic terms for the solvable Lie algebra
gauge of the scalar coset manifold. Finally we will apply our
general results to the case of the $D$-dimensional compactified
$E_{8}\times E_{8}$ heterotic string to derive its bosonic matter
field equations which show a degree of complexity due to the
coupling between the coset scalars and the abelian gauge fields.
We will also discuss the relation between different coset
parametrizations. Besides we will reckon the correspondence of the
two equivalent formulations of the symmetric space sigma model
with couplings which are based on different scalar coset
representations.

In section two we give a brief review of the Kaluza-Klein
reduction of the ten dimensional $\mathcal{N=}1$ supergravity that
is coupled to the abelian Yang-Mills theory \cite{d=10,tani15}
which is studied in detail in \cite{heterotic}. The main objective
of section two will be the construction of the $D$-dimensional
bosonic lagrangian in a compact form by using the solvable Lie
algebra gauge. In section three again under the solvable Lie
algebra gauge we will derive the field equations of the symmetric
space sigma model which is generally coupled to a dilaton, abelian
gauge fields and a Chern-Simons type field strength. In addition
we will mention about three examples on which the results can be
applied. In the last section we will write down the bosnonic
matter field equations of the $D$-dimensional toroidally
compactified heterotic string by using the results obtained in
section three. We will also discuss the possible field
transformations between the two equivalent formulations of the
bosonic lagrangian of the compactified heterotic string.
\section{Toroidally Compactified Heterotic String}\label{section23}
In this section we will focus on the Kaluza-Klein reduction on the
Euclidean torus $T^{10-D}$ of the bosonic sector of the ten
dimensional simple $\mathcal{N=}1$ supergravity \cite{d=10,tani15}
which is coupled to $N$ abelian gauge multiplets. We will not
cover the details of the reduction steps. A detailed study of the
reduction can be referred in \cite{heterotic} in which the
structure of the scalar cosets and the matter couplings which
appear in each dimension upon the dimensional reduction of the
basic fields of the ten dimensional theory are given in detail.
When as a special case, the number of the $U(1)$ gauge fields is
chosen to be $16$, the ten dimensional supergravity which is
coupled to $16$ abelian $U(1)$ gauge multiplets becomes the low
energy effective limit of the $E_{8}\times E_{8}$ ten dimensional
heterotic string theory which forms the massless background
coupling \cite{kiritsis}. Thus when $N=16$ our formulation
corresponds to the dimensional reduction of the low energy
effective bosonic lagrangian of the ten dimensional $E_{8}\times
E_{8}$ heterotic superstring theory. For $N=16$, the $D=10$
Yang-Mills supergravity \cite{d=10,tani15} has the $E_{8}\times
E_{8}$ Yang-Mills gauge symmetry, however the general Higgs vacuum
structure causes a spontaneous symmetry breakdown so that the full
symmetry $E_{8}\times E_{8}$ can be broken down to its maximal
torus subgroup $U(1)^{16}$, whose Lie algebra is the Cartan
subalgebra of $E_{8}\times E_{8}$. Thus the ten dimensional
Yang-Mills supergravity reduces to its maximal torus subtheory
which is an abelian (Maxwell-Einstein) supergravity theory. The
bosonic sector of this abelian Yang-Mills supergravity corresponds
to the low energy effective theory of the bosonic sector of the
fully Higgsed ten dimensional $E_{8}\times E_{8}$ heterotic string
theory \cite{kiritsis,heterotic}. Therefore our main concern will
be the abelian Yang-Mills supergravity in ten dimensions, however
for the purpose of generality we will consider the coupling of $N$
$U(1)$ gauge field multiplets to the graviton multiplet.

As we have discussed before the scalar sectors of the lower
dimensional theories can be formulated as non-linear sigma models,
more specifically as the symmetric space sigma models. One needs
to apply the method of dualisation for some of the fields in
certain dimensions to construct the scalar cosets as symmetric
space sigma models. The scalar cosets $G/K$ are based on the
global internal symmetry groups $G$ which are in general
non-compact real forms of semi-simple Lie groups. In supergravity
theories the global symmetry of the scalar lagrangian can be
extended to the entire bosonic sector of the theory. Under certain
conditions the global internal symmetry groups may be maximally
non-compact (split) real forms but in general they are elements of
a bigger class of Lie groups which contains the global internal
symmetry groups of the maximal supergravities namely the split
real forms as a special subset. The main difference between the
scalar cosets based on the non-compact and the maximally
non-compact numerator groups is the parametrization one can choose
for the coset representatives. For the general non-compact real
forms one can make use of the solvable Lie algebra gauge
\cite{fre} to parameterize the scalar coset. The solvable Lie
algebra is a subalgebra of the Borel algebra of $g$ (the Lie
algebra of $G$). It is simply composed of certain Cartan and
positive root generators of $g$ \cite{hel}.

The bosonic lagrangian of the $D=10$, $\mathcal{N}=1$ abelian
Yang-Mills supergravity which is coupled to $N$ $U(1)$ gauge
multiplets is \cite{d=10,tani15,heterotic}
\begin{equation}\label{ch231}
{\mathcal{L}}_{10}=R\ast1-\frac{1}{2}\ast d\phi_{1}\wedge
d\phi_{1}-\frac{1}{2}e^{\phi_{1}}\ast F_{(3)}\wedge F_{(3)}
-\frac{1}{2}e^{\frac{1}{2}\phi_{1}}\sum\limits_{I=1}^{N}\ast
G_{(2)}^{I}\wedge G_{(2)}^{I},
\end{equation}
where $G_{(2)}^{I}=dB_{(1)}^{I}$ are the $N$ $U(1)$ gauge field
strengths. In \eqref{ch231} $\phi_{1}$ is a scalar field and the
subscripts for the rest of the fields and the field strengths
denote the degree of the fields. We also define the field strength
of the field $A_{(2)}$ as
\begin{equation}\label{ch232}
F_{(3)}=dA_{(2)}+\frac{1}{2}B_{(1)}^{I}\wedge d B_{(1)}^{I}.
\end{equation}
The pure supergravity sector of \eqref{ch231} can be truncated
from the IIA supergravity \cite{2A1,2A2,2A3} by choosing the extra
$R$-$R$ fields to be zero in the IIA lagrangian. The Kaluza-Klein
ansatz for the ten dimensional spacetime metric upon
$T^{n}$-reduction where $n=10-D$ is \cite{nej112}
\begin{equation}\label{ch233}
ds_{10}^{2}=e^{\overset{\rightarrow}{s}\cdot
\overset{\rightarrow}{\phi^{\prime}}}ds_{D}^{2}+\sum\limits_{i=2}^{11-D}
e^{2\overset{\rightarrow}{\gamma}_{i} \cdot
\overset{\rightarrow}{\phi^{\prime}}}(h^{i})^{2}.
\end{equation}
Here we have
\begin{subequations}\label{ch234}
\begin{align}
\overset{\rightarrow}{s}&=(s_{2},s_{3},...,s_{(11-D)}),\notag\\
\notag\\
\overset{\rightarrow}{f}_{i}&=(0,...,0,(10-i)s_{i},s_{i+1},...,s_{(11-D)}),\tag{\ref{ch234}}
\end{align}
\end{subequations}
where in the second line there are $i-2$ zeros and
\begin{equation}\label{yeni1}
s_{i}=\sqrt{\frac{2}{((10-i)(9-i))}}.
\end{equation}
Also we define
\begin{equation}\label{yeni2}
\overset{\rightarrow}{\gamma}_{i}=\frac{1}{2}
\overset{\rightarrow}{s}-\frac{1}{2}\overset{\rightarrow}{f}_{i},
\end{equation}
and $h^{i}$ is
\begin{equation}\label{ch227}
h^{i}=dz^{i}+{\mathcal{A}}_{(1)}^{i}+{\mathcal{A}}_{(0)j}^{i}dz^{j},
\end{equation}
where $\mathcal{A}_{(1)}^{i}$ are the $D$-dimensional Maxwell
gauge fields and $\mathcal{A}_{(0)j}^{i}$ are the $D$-dimensional
scalars. The coordinates $\{z^{i}\}$ are the coordinates on the
$n$-torus $T^{n}$. We further define the vector
\begin{equation}\label{yeni3}
\overset{\rightarrow}{\phi^{\prime}}=(\phi_{2},\phi_{3},...,\phi_{(11-D)}),
\end{equation}
whose components are the dilatons of the Kaluza-Klein reduction.

We should also mention how to built up an ansatz to reduce a
general $D$-dimensional $(n-1)$-form potential field
$A^{D}_{(n-1)}$ on $S^{1}$, with the assumption
 that the $D$-dimensional spacetime is composed of the Cartesian product of
 a $(D-1)$-dimensional subspacetime and $S^{1}$. The ansatz can be
 chosen as \cite{pope}
\begin{equation}\label{ch2129}
A^{D}_{(n-1)}(x,z)=A_{(n-1)}(x)+A_{(n-2)}(x)\wedge dz,
\end{equation}
 where the coordinates $x$ are on the $(D-1)$-dimensional
 spacetime and $z$ is the coordinate on
$S^{1}$. When we take the exterior derivative of \eqref{ch2129}
 we get
\begin{subequations}\label{ch2130}
\begin{align}
F^{D}_{(n)}(x,z)&=dA^{D}_{(n-1)}(x,z)\notag\\
\notag\\
&=dA_{(n-1)}(x)+dA_{(n-2)}(x)\wedge dz.\tag{\ref{ch2130}}
\end{align}
\end{subequations}
While choosing the $(D-1)$-dimensional field strength of
$A_{(n-2)}$ as $F_{(n-1)}=dA_{(n-2)}$, one does not simply choose
the $(D-1)$-dimensional field strength of $A_{(n-1)}$ as
$F_{(n)}=dA_{(n-1)}$ but for the purpose of obtaining a nice
looking lower dimensional lagrangian one defines the lower
dimensional field strengths through the transgression relations
\begin{subequations}\label{ch2131}
\begin{align}
F_{(n)}(x)&=dA_{(n-1)}(x)-dA_{(n-2)}(x)\wedge{\mathcal{A}}(x),\notag\\
\notag\\
F_{(n-1)}(x)&=dA_{(n-2)}(x).\tag{\ref{ch2131}}
\end{align}
\end{subequations}
In terms of these $(D-1)$-dimensional field strengths the
$D$-dimensional field strength can be given as
\begin{equation}\label{ch2132}
F^{D}_{(n)}(x,z)=F_{(n)}(x)+F_{(n-1)}(x)\wedge(dz+{\mathcal{A}}(x)).
\end{equation}
We can express the $D$-dimensional kinetic term of $F_{(n)}^{D}$
in terms of the $(D-1)$-dimensional field strengths we have
defined as
\begin{subequations}\label{ch2133}
\begin{align}
{\mathcal{L}}_{F}^{D}&=-\frac{1}{2}\ast F^{D}_{(n)}\wedge F^{D}_{(n)}\notag\\
\notag\\
&=(-\frac{1}{2}e^{-2(n-1)\alpha\phi}\ast^{(D-1)}F_{(n)}\wedge
F_{(n)}\notag\\
\notag\\
&\quad
-\frac{1}{2}e^{2(D-n-1)\alpha\phi}\ast^{(D-1)}F_{(n-1)}\wedge
F_{(n-1)})\wedge dz\notag\\
\notag\\
&={\mathcal{L}}_{F}^{(D-1)}\wedge dz.\tag{\ref{ch2133}}
\end{align}
\end{subequations}
If we perform the $S^{1}$-reduction step by step on the ten
dimensional metric by using the ansatz \eqref{ch233} and on the
other fields in \eqref{ch231} as well as on the by-product lower
dimensional fields by using the ansatz \eqref{ch2129} then the ten
dimensional lagrangian \eqref{ch231} can be written as
\begin{equation}\label{ch235}
{\mathcal{L}}_{10}={\mathcal{L}}_{D}\wedge
dz^{1}\wedge\cdot\cdot\cdot\wedge dz^{n},
\end{equation}
where the $S^{1}$-reduction is performed $n$ times. The
$D$-dimensional lagrangian in \eqref{ch235} can be calculated in
terms of the $D$-dimensional fields as
\begin{subequations}\label{ch236}
\begin{align}
{\mathcal{L}}_{D}&=R\ast1-\frac{1}{2}\ast
d\overset{\rightarrow}{\phi}\wedge
d\overset{\rightarrow}{\phi}-\frac{1}{2}e^{\overset{\rightarrow}{a}_{1}\cdot
\overset{\rightarrow}{\phi}}\ast F_{(3)}\wedge F_{(3)}\notag\\
\notag\\
&\quad
-\frac{1}{2}\sum\limits_{i}^{}e^{\overset{\rightarrow}{a}_{1i}\cdot
\overset{\rightarrow}{\phi}}\ast F_{(2)i}\wedge F_{(2)i}
-\frac{1}{2}\sum\limits_{i<j}^{}e^{\overset{\rightarrow}{a}_{1ij}\cdot
\overset{\rightarrow}{\phi}}\ast F_{(1)ij}\wedge F_{(1)ij}\notag\\
\notag\\
&\quad
-\frac{1}{2}\sum\limits_{I}^{}e^{\overset{\rightarrow}{c}\cdot
\overset{\rightarrow}{\phi}}\ast G_{(2)}^{I}\wedge
G_{(2)}^{I}-\frac{1}{2}\sum\limits_{i,I}^{}e^{\overset{\rightarrow}{c}_{i}\cdot
\overset{\rightarrow}{\phi}}\ast G_{(1)i}^{I}\wedge G_{(1)i}^{I}\notag\\
\notag\\
&\quad
-\frac{1}{2}\sum\limits_{i}^{}e^{\overset{\rightarrow}{b}_{i}\cdot
\overset{\rightarrow}{\phi}}\ast {\mathcal{F}}_{(2)}^{i}\wedge
{\mathcal{F}}_{(2)}^{i}
-\frac{1}{2}\sum\limits_{i<j}^{}e^{\overset{\rightarrow}{b}_{ij}\cdot
\overset{\rightarrow}{\phi}}\ast {\mathcal{F}}_{(1)j}^{i}\wedge
{\mathcal{F}}_{(1)j}^{i},\tag{\ref{ch236}}
\end{align}
\end{subequations}
where $i,j=2,...,11-D$ and
$\overset{\rightarrow}{\phi}=(\phi_{1},\phi_{2},...,\phi_{(11-D)})$
in which except $\phi_{1}$, the rest of the scalars are the
Kaluza-Klein scalars which originate from the ten dimensional
spacetime metric. The transgression relations which define the
field strengths in \eqref{ch236} in terms of the potentials are as
follows
\begin{subequations}\label{ch237}
\begin{align}
F_{(3)}&=dA_{(2)}+\frac{1}{2}B_{(1)}^{I}dB_{(1)}^{I}-(dA_{(1)i}
+\frac{1}{2}B_{(0)i}^{I}dB_{(1)}^{I}+\frac{1}{2}B_{(1)}^{I}dB_{(0)i}^{I})
\gamma_{\;j}^{i}{\mathcal{A}}_{(1)}^{j}\notag\\
\notag\\
&\quad +\frac{1}{2}(dA_{(0)ij}- B_{(0)[i}^{I}dB_{(0)j]}^{I})
\gamma_{\;k}^{i}{\mathcal{A}}_{(1)}^{k}\gamma_{\;m}^{j}{\mathcal{A}}_{(1)}^{m},\notag\\
\notag\\
F_{(2)i}&=\gamma^{k}_{\;i}(dA_{(1)k}+\frac{1}{2}B_{(0)k}^{I}dB_{(1)}^{I}
+\frac{1}{2}B_{(1)}^{I}dB_{(0)k}^{I}\notag\\
\notag\\
&\quad +(dA_{(0)kj}- B_{(0)[k}^{I}dB_{(0)j]}^{I})
\gamma_{\;l}^{j}{\mathcal{A}}_{(1)}^{l}),\notag\\
\notag\\
F_{(1)ij}&=\gamma^{l}_{\;i}\gamma^{m}_{\;j}(dA_{(0)lm}-B_{(0)[l}^{I}dB_{(0)m]}^{I}),\quad
{\mathcal{F}}_{(2)}^{i}=\widetilde{\gamma}^{i}_{\;j}d(\gamma^{j}_{k}{\mathcal{A}}_{(1)}^{k}),\notag\\
\notag\\
G_{(2)}^{I}&=dB_{(1)}^{I}-dB_{(0)i}^{I}\gamma^{i}_{\;j}{\mathcal{A}}_{(1)}^{j},\quad
G_{(1)i}^{I}=\gamma^{j}_{\;i}dB_{(0)j}^{I},\quad
{\mathcal{F}}_{(1)j}^{i}=\gamma_{j}^{k}d{\mathcal{A}}_{(0)k}^{i},\tag{\ref{ch237}}
\end{align}
\end{subequations}
where we have omitted the wedge product. The dilatons
$\overset{\rightarrow}{\phi^{\prime}}$, the Kaluza-Klein-Maxwell
potentials ${\mathcal{A}}_{(1)}^{j}$ as well as the axions
${\mathcal{A}}_{(0)k}^{i}$ in each dimension are the descendants
of the ten dimensional spacetime metric. The potentials
$A_{(0)lm}$, $A_{(1)k}$ and $A_{(2)}$ are the Kaluza-Klein
descendants of the two-form potential in $D=10$. The potentials
$B_{(0)j}^{I}$ and $B_{(1)}^{I}$ are the $D$-dimensional
remainders of the ten dimensional Yang-Mills potentials
$B_{(1)}^{I}$ as a result of the ansatz \eqref{ch2129} applied in
each $S^{1}$-reduction step. We define the matrix
$\gamma_{\;j}^{i}$ as
\begin{equation}\label{ch229}
\gamma_{j}^{i}= [(1+{\mathcal{A}}_{(0)})^{-1}]_{j}^{i},
\end{equation}
and $\widetilde{\gamma}_{\;j}^{i}$ is the inverse of it. The
dilaton vectors which couple to the scalars
$\overset{\rightarrow}{\phi}$ in various field strength terms in
\eqref{ch236} are defined as \cite{heterotic}
\begin{subequations}\label{ch238}
\begin{gather}
\overset{\rightarrow}{a}_{1}=(1,-2\overset{\rightarrow}{s})\quad,\quad
\overset{\rightarrow}{a}_{1i}=(1,\overset{\rightarrow}{f}_{i}-2\overset{\rightarrow}{s})
\quad,\quad
\overset{\rightarrow}{a}_{1ij}=(1,\overset{\rightarrow}{f}_{i}
+\overset{\rightarrow}{f}_{j}-2\overset{\rightarrow}{s}),\notag\\
\notag\\
\overset{\rightarrow}{b}_{i}=(0,-\overset{\rightarrow}{f}_{i})\quad,\quad
\overset{\rightarrow}{b}_{ij}=(0,-\overset{\rightarrow}{f}_{i}+\overset{\rightarrow}{f}_{j}),\notag\\
\notag\\
\overset{\rightarrow}{c}=(\frac{1}{2},-\overset{\rightarrow}{s})\quad,\quad
\overset{\rightarrow}{c}_{i}=
(\frac{1}{2},\overset{\rightarrow}{f}_{i}-\overset{\rightarrow}{s}).\tag{\ref{ch238}}
\end{gather}
\end{subequations}
In \cite{heterotic} the scalar sector and the abelian matter
coupling in each dimension are reformulated through a series of
field and field strength redefinitions. The dualisation of certain
fields is also performed in various dimensions when necessary. In
this way the global symmetry groups are identified. Also the
lagrangian \eqref{ch236} is written in a more compact form whose
scalar sector is formulated as a $G/K$ non-linear sigma model. We
will not mention about the details of this calculation here, we
will only present the result which will be a reference point for
our further analysis in the following sections. In
\cite{heterotic} it is shown that when a single dilaton is
decoupled from the rest of the scalars then the $G/K$ coset
representatives $\nu^{\prime}$ and the internal metric
\begin{equation}\label{29}
\mathcal{M}=\nu ^{\prime T}\nu^{\prime} ,
\end{equation}
generated by the remaining scalars are elements of
$O^{\prime}(10-D+N,10-D)$ $\footnote{The reason why we use a prime
will be clear later. Here we slightly change the notation used in
\cite{heterotic}.}$ which is composed of the
($20-2D+N$)-dimensional real matrices $A$ which satisfy
\begin{equation}\label{29.1}
A^{T}\Omega A=\Omega,
\end{equation}
where the $(20-2D+N)\times(20-2D+N)$ matrix $\Omega$ is
\begin{equation}\label{ch2343}
\Omega=\left(\begin{array}{ccc}
  0 & 0 & -{\mathbf{1}}_{(10-D)} \\
  0 & {\mathbf{1}}_{(N)} & 0 \\
  -{\mathbf{1}}_{(10-D)} & 0 & 0 \\
\end{array}\right),
\end{equation}
in which ${\mathbf{1}}_{(n)}$ is the $n\times n$ unit matrix. The
scalar lagrangian of the $D$- dimensional compactified theories
can be described in the form
\begin{equation}\label{29.5}
 \mathcal{L}_{scalar}=\frac{1}{4}tr(\ast d\mathcal{M}^{-1}\wedge
 d\mathcal{M}),
\end{equation}
with an additional decoupled dilatonic kinetic term after certain
field redefinitions. The coset representatives $\nu^{\prime}$ are
parameterized as
\begin{equation}\label{ch2332}
\nu^{\prime}=e^{\frac{1}{2}\varphi^{i}H_{i}}e^{A^{i}_{(0)j}E_{i}^{j}}e^{\frac{1}{2}A_{(0)ij}
V^{ij}}e^{B^{I}_{(0)i}U_{I}^{i}}.
\end{equation}
Here $\{H_{i},E_{i}^{j},V^{ij},U_{I}^{i}\}$ are
($20-2D+N$)-dimensional matrices and following the notation of
\cite{heterotic} we have modified the ranges of the indices $i,j$
as $i,j=1,...,10-D$. We also assume that $i<j$. The matrices
$\{H_{i},E_{i}^{j},V^{ij},U_{I}^{i}\}$ are \cite{heterotic,sezgin}
\begin{subequations}\label{ch2338}
\begin{gather}
\overset{\rightarrow}{H}_{i}=\left(\begin{array}{ccc}
  \sum_{i}\overset{\rightarrow}{c}_{i}e_{ii}& 0 & 0 \\
  0 & 0 & 0 \\
  0 & 0 & -\sum_{i}\overset{\rightarrow}{c}_{i}e_{ii} \\
\end{array}\right),
\quad E_{i}^{j}=\left(\begin{array}{ccc}
  -e_{ji}& 0 & 0 \\
  0 & 0 & 0 \\
  0 & 0 & e_{ij} \\
\end{array}\right), \notag\\
\notag\\
 V^{ij}=\left(\begin{array}{ccc}
  0& 0 & e_{ij}-e_{ji} \\
  0 & 0 & 0 \\
  0 & 0 &0 \\
\end{array}\right),\quad
U_{I}^{i}=\left(\begin{array}{ccc}
  0& e_{iI} & 0 \\
  0 & 0 & e_{Ii} \\
  0 & 0 & 0 \\
\end{array}\right),\tag{\ref{ch2338}}
\end{gather}
\end{subequations}
where the matrices are partitioned with appropriate dimensions
which can be read from the non-zero entries and $e_{ab}$ is a
matrix with appropriate dimensions and which has zero entries,
except a $\{+1\}$ entry at the $a$'th row and the $b$'th column.
Notice that the indices $i,j=1,...,10-D$ and $I=1,...,N$ determine
the dimensions of the matrices $e_{ab}$. As it is mentioned in
\cite{heterotic} the matrices
$\{H_{i},E_{i}^{j},V^{ij},U_{I}^{i}\}$ form up an algebra and they
can be embedded in a fundamental representation of
$o^{\prime}(10-D+N,10-D)$. As a matter of fact they generate the
solvable Lie algebra of $o^{\prime}(10-D+N,10-D)$ in each
dimension thus the parametrization of the coset in \eqref{ch2332}
is nothing but an example of the solvable Lie algebra gauge. This
result justifies the prediction that the global internal symmetry
group of the scalar lagrangian is $O^{\prime}(10-D+N,10-D)$ and
the scalar manifold for the $D$-dimensional compactified theory
with $N$ gauge multiplet couplings becomes
\begin{subequations}\label{ch2319}
\begin{gather}
\frac{O^{\prime}(10-D+N,10-D)}{O(10-D+N)\times
O(10-D)}\times{\Bbb{R}}.\tag{\ref{ch2319}}
\end{gather}
\end{subequations}
The extra ${\Bbb{R}}$ factor arises since there is an additional
dilaton which is decoupled from the rest of the scalars in the
scalar lagrangian. The group
$O^{\prime}(10-D+N,10-D)\times\mathbb{R}$ is the global internal
symmetry of not only the scalar lagrangian but the entire
$D$-dimensional bosonic lagrangian as well. The lagrangian
\eqref{ch236} upon the above mentioned restoration, in terms of
the newly defined fields can be written as
\begin{subequations}\label{ch2339}
\begin{align}
{\mathcal{L}}_{D}&=R\ast1-\frac{1}{2}\ast d\phi\wedge
d\phi+\frac{1}{4}tr(\ast d{\mathcal{M}}^{-1}\wedge
d{\mathcal{M}})\notag\\
\notag\\
&\quad -\frac{1}{2}e^{-\sqrt{8/(D-2)}\phi}\ast F_{(3)}\wedge
F_{(3)}-\frac{1}{2}e^{-\sqrt{2/(D-2)}\phi}\ast
H_{(2)}^{T}\wedge{\mathcal{M}}H_{(2)},\tag{\ref{ch2339}}
\end{align}
\end{subequations}
where
\begin{equation}\label{yeni4}
H_{(2)}=dC_{(1)},
\end{equation}
in which we define
 \begin{equation}\label{ch2340}
C_{(1)}=\left(\begin{array}{c}
  A_{(1)i} \\
  B_{(1)}^{I} \\
 {\mathcal{A}}_{(1)}^{i} \\
\end{array}\right),
\end{equation}
which is a column vector of dimension $(20-2D+N)$. The field
strength $F_{(3)}$ in \eqref{ch2339} can be given in a compact
form for any dimension as
\begin{equation}\label{ch2342}
F_{(3)}=dA_{(2)}+\frac{1}{2}C_{(1)}^{T}\:\Omega\: dC_{(1)}.
\end{equation}
The definitions of the potentials introduced in \eqref{ch2339} in
terms of the original potentials which come from the Kaluza-Klein
reduction in \eqref{ch236} can be referred in \cite{heterotic}. In
this work, to derive the field equations in the following sections
we will consider the potentials used to construct \eqref{ch2339}
as our starting point. In addition to the general $D$-dimensional
lagrangian \eqref{ch2339} furthermore the $D=4$ and the $D=3$
cases can be studied separately, since they have global symmetry
enhancements over the general scheme of
$O^{\prime}(10-D+N,10-D)\times\mathbb{R}$ \cite{heterotic}. One
may define additional scalars in these dimensions by dualizing
certain fields by applying the lagrange multiplier methods. When
the two-form potential $A_{(2)}$ is dualized with an additional
axion in $D=4$, an axion-dilaton, $SL(2,{\Bbb{R}})$ system is
decoupled from the rest of the scalars in the scalar lagrangian
and the enlarged $D=4$ scalar manifold becomes
\begin{subequations}\label{ch2320}
\begin{gather}
\frac{O^{\prime}(6+N,6)}{O(6)\times
O(6+N)}\times\frac{SL(2,{\Bbb{R}})}{O(2)}.\tag{\ref{ch2320}}
\end{gather}
\end{subequations}
On the other hand in $D=3$, the bosonic fields
$({\mathcal{A}}_{(1)}^{i},A_{(1)i},B_{(1)}^{I})$ can be dualized
to give $7+7+N$ additional axions. In this case the entire bosonic
sector comes out to be composed of only the scalars. The $D=3$
enlarged scalar manifold then becomes
\begin{subequations}\label{ch2321}
\begin{gather}
\frac{O^{\prime}(8+N,8)}{O(8)\times O(8+N)}.\tag{\ref{ch2321}}
\end{gather}
\end{subequations}
We see that all of the global internal symmetry groups in
\eqref{ch2319}, \eqref{ch2320} and \eqref{ch2321} apart from the
contributions of the decoupled scalars namely the groups
$O^{\prime}(10-D+N,10-D)$, $O^{\prime}(6+N,6)$,
$O^{\prime}(8+N,8)$ are non-compact real forms of semi-simple Lie
groups and they enable solvable Lie algebra parametrizations of
the cosets that are generated by the denominator groups
$O(10-D+N)\times O(10-D)$, $O(6)\times O(6+N)$, $O(8)\times
O(8+N)$, respectively.
\section{The General Formalism for the SSSM with Couplings}
In this section we will construct the general formulation of the
abelian gauge matter, dilaton and Chern-Simons couplings of the
generic symmetric space sigma model (SSSM)
\cite{julia1,ker1,ker2,nej1,nej2} under the solvable Lie algebra
parametrization. We will derive the field equations for a general
theory and then we will give a number of examples on which we can
apply the results. Also in the next section we will use the
results to write down the field equations of the $D$-dimensional
low energy effective theory of the $E_{8}\times E_{8}$ heterotic
string which is studied in the previous section. The general
formalism and the field equations are already derived for the
symmetric space sigma model and the abelian matter coupled
symmetric space sigma model in \cite{nej1,nej2} and \cite{nej3}
respectively under the solvable Lie algebra gauge. Therefore in
this section we extend the construction of \cite{nej3} further by
including a dilaton and a Chern-Simons coupling.

We will first construct the scalar lagrangian by using the
solvable Lie algebra parametrization \cite{fre} for the coset
representatives. In the last section we have seen that the
toroidally compactified heterotic string gives the
Maxwell-Einstein supergravities in $D$-dimensions. The scalars
decoupled from a single dilaton are governed by symmetric space
sigma models. The scalar fields parameterize the scalar coset
manifold $O^{\prime}(10-D+N,10-D)/O(10-D+N)\times O(10-D)$ where
$O^{\prime}(10-D+N,10-D)$ is in general a non-compact real form of
a semi-simple Lie group and $O(10-D+N)\times O(10-D)$ is its
maximal compact subgroup. For this reason the scalar manifold is a
Riemannian globally symmetric space for all the
$O^{\prime}(10-D+N,10-D)$-invariant Riemannian structures on it
\cite{hel}.

In this section we will use a slightly different notation. We will
take the representation of the global symmetry group as
$O(10-D,10-D+N)$ whose ($20-2D+N$)-dimensional real matrix
elements $A$ satisfy the defining relation
\begin{equation}\label{24.1}
A^{T}\eta A=\eta,
\end{equation}
where
\begin{equation}\label{24.5}
\eta=\text{diag}(-,-,...,-,+,+,...,+),
\end{equation}
is the indefinite signature metric with $10-D$ minus signs and
$10-D+N$ plus signs. Although the set of matrices defined in
\eqref{29.1} and \eqref{24.1} may differ the groups they form are
isomorphic to each other \cite{knapp}. Thus in this section we
will base our formulation on the scalar coset which is the
conventional one used in the construction of the Maxwell-Einstein
supergravities in \cite{d=7,d=8,d=9}.

To construct the symmetric space sigma model lagrangian for the
scalar coset manifold,
\begin{equation}\label{yeni5}
\frac{O(10-D,10-D+N)}{O(10-D)\times O(10-D+N)},
\end{equation}
one may make use of the solvable Lie algebra parametrization
\cite{fre} for the parametrization of the coset representatives.
The solvable Lie algebra gauge or the parametrization is a
consequence of the Iwasawa decomposition \cite{hel}
\begin{equation}\label{21}
\begin{aligned}
o(10-D,10-D+N)&=\mathbf{k}_{0}\oplus \mathbf{s}_{0}\\
\\
&=\mathbf{k}_{0}\oplus \mathbf{h_{k}}\oplus
\mathbf{n_{k}},
\end{aligned}
\end{equation}
where $\mathbf{k}_{0}$ is the Lie algebra of $O(10-D)\times
O(10-D+N)$ which is a maximal compact subgroup of $O(10-D,10-D+N)$
and $\mathbf{s}_{0}$ is a solvable Lie subalgebra of
$o(10-D,10-D+N)$. In \eqref{21} $\mathbf{h}_{k}$ is a subalgebra
of the Cartan subalgebra $\mathbf{h}_{0}$ of $o(10-D,10-D+N)$
which generates the maximal R-split torus in $O(10-D,10-D+N)$
\cite{hel,ker2,nej2}. The nilpotent Lie subalgebra
$\mathbf{n_{k}}$ of $o(10-D,10-D+N)$ is generated by a subset
$\{E_{m}\}$ of the positive root generators of $o(10-D,10-D+N)$
where $m\in\Delta_{nc}^{+}$. The roots in $\Delta_{nc}^{+}$ are
the non-compact roots with respect to the Cartan involution
$\theta$ induced by the Cartan decomposition \cite{hel,nej3,nej2}
\begin{equation}\label{22}
o(10-D,10-D+N)=\mathbf{k}_{0}\oplus\mathbf{u}_{0},
\end{equation}
where $\mathbf{u}_{0}$ is a vector subspace of $o(10-D,10-D+N)$.
We should remark that when the Lie group $O(10-D,10-D+N)$ is in
split real form (maximally non-compact) \cite{hel} then for a
Cartan decomposition and a resulting Iwasawa decomposition the
solvable Lie algebra $\mathbf{s}_{0}$ in \eqref{21} coincides with
the Borel subalgebra which is generated by all the simple root
Cartan generators $\{H_{\alpha}\}$ (or any other basis) of
$\mathbf{h_{0}}$ and the positive root generators
$\{E_{\alpha}\mid\alpha\in \Delta^{+}\}$ \cite{julia1}. The split
real form global symmetry groups are of the form $O(n,n)$,
 $O(n+1,n)$, $O(n,n+1)$ \cite{pope}.

Since we have \cite{pope}
\begin{equation}\label{22.5}
\text{dim}\mathbf{s}_{0}=(10-D)\times(10-D+N),
\end{equation}
by coupling $(10-D)\times(10-D+N)$ scalars to the generators of
the solvable Lie algebra $\mathbf{s}_{0}$ we can parameterize the
representatives of the scalar coset manifold in the solvable Lie
algebra gauge as \cite{hel}
\begin{equation}\label{23}
\nu=e^{\frac{1}{2}\phi ^{i}H_{i}}e^{\chi ^{m}E_{m}},
\end{equation}
where $\{H_{i}\}$ for $i=1,...,$ dim($\mathbf{h}_{k})\equiv r$ are
the generators of $\mathbf{h}_{k}$ and $\{E_{m}\}$ for
$m\in\Delta_{nc}^{+}$ are the positive root generators which
generate $\mathbf{n_{k}}$. The scalars $\{\phi^{i}\}$ for
$i=1,...,r$ are called the dilatons and $\{\chi^{m}\}$ for
$m\in\Delta_{nc}^{+}$ are called the axions. The coset
representatives $\nu$ satisfy the defining relation of
$O(10-D,10-D+N)$ namely
\begin{equation}\label{24}
\nu ^{T}\eta \nu=\eta.
\end{equation}
If we assume that we choose the ($20-2D+N$)-dimensional
fundamental representation for the algebra $o(10-D,10-D+N)$ and if
we define the internal metric
\begin{equation}\label{25}
\mathcal{M}=\nu ^{T}\nu,
\end{equation}
then the scalar lagrangian which governs the $(10-D)\times
(10-D+N)$ scalar fields of the theory can be constructed as
\begin{equation}\label{26}
 \mathcal{L}_{scalar}=\frac{1}{4}tr(\ast d\mathcal{M}^{-1}\wedge
 d\mathcal{M}).
\end{equation}
Since in this work our main perspective is the derivation of the
field equations of the toroidally compactified heterotic string we
take the scalar coset as $O(10-D,10-D+N)/O(10-D)\times O(10-D+N)$
however it is obvious that our analysis is valid for any scalar
coset of the form $O(p,q)/O(p)\times O(q)$. As we have assumed a
($20-2D+N$)-dimensional representation for the Lie algebra
$o(10-D,10-D+N)$ we can couple ($20-2D+N$) abelian gauge fields
$A^{I}$ and a dilaton $\sigma$ to the scalars as
\cite{nej3,ker1,ker2,nej1,nej2}
\begin{subequations}\label{211.5}
\begin{align}
\mathcal{L}_{m}&=c_{2}e^{\alpha_{1}\sigma}\ast
F\wedge\mathcal{M}F \notag\\
\notag\\
\quad &=c_{2}e^{\alpha_{1}\sigma}\mathcal{M}_{IJ} \ast F^{I}\wedge
F^{J},\tag{\ref{211.5}}
\end{align}
\end{subequations}
where the field strengths of the gauge fields are defined to be
$F^{I}=dA^{I}$. Furthermore one can couple a two-form field $B$ to
the other field content by introducing the Chern-Simons three-form
\begin{equation}\label{212}
 G=dB+c_{4}\eta_{IJ}A^{I}\wedge F^{J},
\end{equation}
which becomes the field strength of $B$ indeed. We are ready to
write down the general lagrangian of the $O(p,q)/O(p)\times O(q)$
scalar coset which has the abelian gauge, Chern-Simons and the
dilaton couplings in $D$-dimensional spacetime,
\begin{equation}\label{210}
\begin{aligned}
\mathcal{L}&=c_{1}\ast d\sigma\wedge d\sigma
+c_{2}e^{\alpha_{1}\sigma} \ast F\wedge\mathcal{M}
F\\
\\
&\quad +\frac{1}{4}tr( \ast d\mathcal{M}^{-1}\wedge
d\mathcal{M})+c_{3}e^{\alpha_{2}\sigma}\ast G \wedge G.
\end{aligned}
\end{equation}
Our field content becomes $\{\sigma,A^{I},B,\phi^{j},\chi^{m}\}$
in which we have $(10-D)\times(10-D+N)$ scalars of the coset,
($20-2D+N$) abelian gauge fields $A^{I}$, a two-form $B$ and a
single dilaton $\sigma$ which we take to be decoupled from the
rest of the scalars that parameterize the scalar coset manifold.

We will first vary the lagrangian \eqref{210} with respect to the
dilaton $\sigma$. The Euler-Lagrange equation \cite{thiring} which
yields the vanishing of the variation of \eqref{210} with respect
to $\sigma$ gives
\begin{equation}\label{211.6}
\begin{aligned}
(-1)^{(D-1)}d(2c_{1}\ast
d\sigma)=&c_{3}\alpha_{2}e^{\alpha_{2}\sigma} \ast G\wedge
G\\
\\
&+c_{2}\alpha_{1}e^{\alpha_{1}\sigma}\mathcal{M}_{IJ} \ast
F^{I}\wedge F^{J}.
\end{aligned}
\end{equation}
The left hand side comes from the variation of the kinetic term of
the dilaton in \eqref{210} and the terms in the right hand side
arise due to the coupling of the dilaton to the other fields.
Since the lagrangian \eqref{210} does not explicitly depend on $B$
the variation of it with respect to $B$ will yield the
Euler-Lagrange equation which defines a closed form. It is

\begin{equation}\label{211.7}
-(-1)^{(D+1)}d(c_{3}e^{\alpha_{2}\sigma}(2\ast dB+ 2c_{4}\eta_{IJ}
\ast (A^{I}\wedge F^{J})))=0.
\end{equation}
By using the definition of the Chern-Simons three-form given in
\eqref{212} we can write \eqref{211.7} as
\begin{equation}\label{211.8}
d(e^{\alpha_{2}\sigma}\ast G)=0.
\end{equation}
We will now vary the lagrangian \eqref{210} with respect to the
abelian gauge fields $A^{I}$. The Euler-Lagrange equations of
motion read
\begin{equation}\label{211.9}
\begin{aligned}
0=&(-1)^{3(D-3)}2c_{3}c_{4}e^{\alpha_{2}\sigma}\eta_{IK}
F^{K}\wedge(\ast dB+c_{4}\eta_{LM}\ast(A^{L}\wedge F^{M}))\\
\\
&+d(c_{2}e^{\alpha_{1}\sigma}(\mathcal{M}_{IJ}\ast
dA^{J}+\mathcal{M}_{JI}\ast dA^{J})\\
\\
&+ 2c_{3}c_{4}e^{\alpha_{2}\sigma}(\ast
dB+c_{4}\eta_{LM}\ast(A^{L}\wedge F^{M}))\wedge \eta_{JI} A^{J}).
\end{aligned}
\end{equation}
Bearing in mind that from its definition in \eqref{25} the
internal metric $\mathcal{M}$ is symmetric so that
\begin{equation}\label{280}
\mathcal{M}_{IJ}=\mathcal{M}_{JI},
\end{equation}
also by using the definition of the Chern-Simons form in
\eqref{212} the Euler-Lagrange equations \eqref{211.9} for the
gauge fields $A^{I}$ can be written as
\begin{equation}\label{281}
\begin{aligned}
(-1)^{3(D-3)}c_{3}c_{4}e^{\alpha_{2}\sigma}\eta_{IK}
F^{K}\wedge\ast G
 =&-d(c_{2}e^{\alpha_{1}\sigma}\mathcal{M}_{IJ}\ast dA^{J})\\
 \\
 &-d(c_{3}c_{4}e^{\alpha_{2}\sigma}\ast G\wedge\eta_{JI} A^{J}).
\end{aligned}
\end{equation}
Now by acting the exterior derivative the last term in \eqref{281}
can be written as
\begin{equation}\label{282}
\begin{aligned}
-d(c_{3}c_{4}e^{\alpha_{2}\sigma}\ast G\wedge\eta_{JI}
A^{J})=&-c_{3}c_{4}(d(e^{\alpha_{2}\sigma}\ast G)\wedge\eta_{JI}
A^{J}\\
\\
&+(-1)^{(D-3)}e^{\alpha_{2}\sigma}\ast G\wedge\eta_{JI} dA^{J}).
\end{aligned}
\end{equation}
If we use the field equation \eqref{211.8} of the two-form field
$B$ above and insert the result back in \eqref{281} we obtain
\begin{equation}\label{282.5}
\begin{aligned}
(-1)^{3(D-3)}c_{3}c_{4}e^{\alpha_{2}\sigma}\eta_{IK}
F^{K}\wedge\ast G
 =&-d(c_{2}e^{\alpha_{1}\sigma}\mathcal{M}_{IJ}\ast dA^{J})\\
 \\
 &-(-1)^{(D-3)}c_{3}c_{4}e^{\alpha_{2}\sigma}\ast G\wedge\eta_{JI} dA^{J}.
\end{aligned}
\end{equation}
Noting that $\eta$ is symmetric,
\begin{equation}\label{282.7}
\eta_{IJ}=\eta_{JI},
\end{equation}
we can finally write the field equations of $A^{I}$ as
\begin{equation}\label{283}
d(c_{2}e^{\alpha_{1}\sigma}\mathcal{M}_{IJ}\ast
F^{J})=(-1)^{D}2c_{3}c_{4}e^{\alpha_{2}\sigma}\eta_{IJ}F^{J}\wedge\ast
G,
\end{equation}
which we will compactly express in matrix form as
\begin{equation}\label{284}
d(c_{2}e^{\alpha_{1}\sigma}\mathcal{M}\ast
F)=(-1)^{D}2c_{3}c_{4}e^{\alpha_{2}\sigma}\eta F\wedge\ast G.
\end{equation}
One can take the internal metric out of the exterior derivative on
the left hand side above to obtain
\begin{equation}\label{285}
\begin{aligned}
d(c_{2}e^{\alpha_{1}\sigma}\ast
F)=&-c_{2}e^{\alpha_{1}\sigma}\mathcal{M}^{-1}d\mathcal{M}\wedge\ast
F\\
\\
&+(-1)^{D}2c_{3}c_{4}e^{\alpha_{2}\sigma}\mathcal{M}^{-1}\eta
F\wedge\ast G.
\end{aligned}
\end{equation}
We may also express the field equations \eqref{283} in terms of
the Cartan-Maurer form
\begin{equation}\label{286}
\mathcal{G}=d\nu\nu^{-1}.
\end{equation}
Before going in that direction we should remark on certain
identities which the general coset representatives \eqref{23}
satisfy. First of all since
\begin{equation}\label{287}
\nu\nu^{-1}=\nu^{-1}\nu=\mathbf{1},
\end{equation}
taking the exterior derivative of both sides yields the identities
\begin{equation}\label{288}
d\nu\nu^{-1}=-\nu d\nu^{-1},\quad d\nu^{-1}\nu=-\nu^{-1}d\nu.
\end{equation}
Also as $\eta$ is a symmetric matrix and
\begin{equation}\label{288.5}
\eta^{2}=\mathbf{1},
\end{equation}
starting from \eqref{24} one can derive the following identities
\begin{subequations}
\begin{gather}\label{289}
\eta\nu^{T}\eta=\nu^{-1},\quad\eta\nu^{-1}\eta=\nu^{T},\quad\nu\eta\nu^{T}=\eta,\notag\\
 \notag\\
 \nu^{-1}\eta(\nu^{T})^{-1}=\eta,\quad\eta\nu\eta=(\nu^{T})^{-1},\quad\eta(\nu^{T})^{-1}\eta=\nu\tag{\ref{289}}.
\end{gather}
\end{subequations}
Since $\eta$ is a constant matrix one can take the exterior
derivative of the above identities to obtain
\begin{equation}\label{289.4}
\eta d\nu^{T}\eta=d\nu^{-1},\quad\eta
d\nu^{-1}\eta=d\nu^{T},\quad\eta d\nu\eta=d(\nu^{T})^{-1}.
\end{equation}
Due to the definition of the coset elements in \eqref{23} we also
have
\begin{equation}\label{289.5}
(\nu^{T})^{-1}=(\nu^{-1})^{T}.
\end{equation}
In order to write the field equations of $A^{I}$ in terms of the
Cartan-Maurer form $\mathcal{G}$ starting from \eqref{284} we can
proceed as follows
\begin{equation}\label{289.6}
\begin{aligned}
d(c_{2}e^{\alpha_{1}\sigma}\mathcal{M}\ast
F)=&c_{2}e^{\alpha_{1}\sigma}d\nu^{T}\wedge\nu\ast
F\\
\\
&+\nu^{T}(c_{2}e^{\alpha_{1}\sigma}d\nu\wedge \ast F+\nu
d(c_{2}e^{\alpha_{1}\sigma}\ast F)).
\end{aligned}
\end{equation}
Inserting this result back in \eqref{284} we obtain
\begin{equation}\label{290}
\begin{aligned}
d(c_{2}e^{\alpha_{1}\sigma}\ast
F)=&-c_{2}e^{\alpha_{1}\sigma}\nu^{-1}d\nu\wedge \ast
F\\
\\
&-c_{2}e^{\alpha_{1}\sigma}\nu^{-1}(d\nu\nu^{-1})^{T}\nu\wedge\ast
F\\
\\
&+(-1)^{D}2c_{3}c_{4}e^{\alpha_{2}\sigma}\nu^{-1}(\nu^{T})^{-1}\eta
F\wedge\ast G.
\end{aligned}
\end{equation}
Now multiplying the above equation with $\eta$ on both sides and
then using the identities \eqref{288}-\eqref{289.5} effectively we
obtain
\begin{equation}\label{291}
\begin{aligned}
d(c_{2}e^{\alpha_{1}\sigma}\eta\ast
F)=&c_{2}e^{\alpha_{1}\sigma}d\nu^{T}(\nu^{T})^{-1}\wedge \eta\ast
F\\
\\
&+c_{2}e^{\alpha_{1}\sigma}\nu^{T}(d\nu\nu^{-1})(\nu^{T})^{-1}\wedge\eta\ast
F\\
\\
&+(-1)^{D}2c_{3}c_{4}e^{\alpha_{2}\sigma}\mathcal{M} F\wedge\ast
G.
\end{aligned}
\end{equation}
By using
\begin{equation}\label{292}
d\nu^{T}(\nu^{T})^{-1}=\nu^{T}\mathcal{G}^{T}(\nu^{T})^{-1},
\end{equation}
finally we can express \eqref{284} in terms of the Cartan-Maurer
form \eqref{286} as
\begin{equation}\label{293}
\begin{aligned}
d(c_{2}e^{\alpha_{1}\sigma}\eta\ast
F)=&c_{2}e^{\alpha_{1}\sigma}(\nu^{T}(\mathcal{G}^{T}+\mathcal{G})(\nu^{T})^{-1})\wedge
\eta\ast
F\\
\\
&+(-1)^{D}2c_{3}c_{4}e^{\alpha_{2}\sigma}\mathcal{M} F\wedge\ast
G.
\end{aligned}
\end{equation}
For the generic non-split scalar cosets the Cartan-Maurer form
$\mathcal{G}$ is calculated explicitly in terms of the coset
scalars namely the dilatons $\{\phi^{i}\}$ and the axions
$\{\chi^{m}\}$ in \cite{nej2}. It reads
\begin{equation}\label{294.6}
\begin{aligned}
\mathcal{G}&=\frac{1}{2}d\phi ^{i}H_{i}+e^{%
\frac{1}{2}\beta _{i}\phi ^{i}}U^{\beta }E_{\beta }\\
\\
&=\frac{1}{2}d\phi ^{i}H_{i}+\overset{\rightharpoonup }{
\mathbf{E}^{\prime }}\:\mathbf{\Omega }\:\overset{\rightharpoonup
}{d\chi },
\end{aligned}
\end{equation}
where $\{H_{i}\}$ for $i=1,...,r$ are the generators of
$\mathbf{h}_{k}$ and $\{E_{\beta}\}$ for $\beta\in\Delta_{nc}^{+}$
are the generators of $\mathbf{n}_{k}$ as we have already defined
before. We have introduced the column vector
\begin{equation}\label{295}
U^{\alpha}=\mathbf{\Omega}^{\alpha}_{\beta}d\chi^{\beta},
\end{equation}
and the row vector
\begin{equation}\label{296}
(\overset{\rightharpoonup }{
\mathbf{E}^{\prime }})_{\beta}=e^{%
\frac{1}{2}\beta _{i}\phi ^{i}}E_{\beta},
\end{equation}
where the components of the roots $\beta\in\Delta_{nc}^{+}$ are
defined as
\begin{equation}\label{297}
[H_{i},E_{\beta}]=\beta_{i}E_{\beta}.
\end{equation}
Also $\mathbf{\Omega}$ is a
dim$\mathbf{n_{k}}\times$dim$\mathbf{n_{k}}$ matrix
\begin{equation}\label{298}
\begin{aligned}
 \mathbf{\Omega}&=\sum\limits_{m=0}^{\infty }\dfrac{\omega
^{m}}{(m+1)!}\\
\\
&=(e^{\omega}-I)\,\omega^{-1}.
\end{aligned}
\end{equation}
The components of the matrix $\omega$ are
\begin{equation}\label{299}
\omega_{\beta}^{\gamma}=\chi^{\alpha}K_{\alpha\beta}^{\gamma}.
\end{equation}
The structure constants $K_{\alpha\beta}^{\gamma}$ are defined as
\begin{equation}\label{300}
[E_{\alpha},E_{\beta}]=K_{\alpha\beta}^{\gamma}E_{\gamma}.
\end{equation}
In other words since
\begin{equation}\label{301}
[E_ {\alpha},E_{\beta}]=N_{\alpha,\beta}E_{\alpha+\beta},
\end{equation}
we have
\begin{subequations}\label{yeni6}
\begin{gather}
K_{\beta\gamma}^{\alpha}=N_{\beta,\gamma} \quad\text{if}\quad
\beta+\gamma=\alpha\quad ,\quad K_{\beta\gamma}^{\alpha}=0
\quad\text{if}\quad
\beta+\gamma\neq\alpha,\notag\\
\notag\\
K_{\beta\beta}^{\alpha}=0.\tag{\ref{yeni6}}
\end{gather}
\end{subequations}
The reason why we have written the field equations of the gauge
fields $A^{I}$ \eqref{284} in terms of the Cartan-Maurer form
$\mathcal{G}$ in \eqref{293} becomes clear now. After choosing a
($20-2D+N$)-dimensional representation for the Lie algebra
$o(10-D,10-D+N)$ (thus for the solvable Lie subalgebra
$\mathbf{s}_{0}$) and after expressing the coset representatives
\eqref{23} and the internal metric \eqref{25} as matrices one can
insert \eqref{294.6} in \eqref{293} to write the later explicitly
in terms of the coset scalars $\phi^{i}$ and $\chi^{m}$.

The terms in the lagrangian \eqref{210} which contain the coset
scalars $\phi^{i}$ and $\chi^{m}$ are
\begin{equation}\label{302}
\mathcal{L}(\phi^{i},\chi^{m})=\frac{1}{4}tr( \ast
d\mathcal{M}^{-1}\wedge d\mathcal{M})+c_{2}e^{\alpha_{1}\sigma}
\ast F\wedge\mathcal{M} F.
\end{equation}
By using the results of \cite{ker1,nej1,nej2} we can write
\eqref{302} as
\begin{equation}\label{303}
\mathcal{L}(\phi^{i},\chi^{m})=-\frac{1}{2}tr(\ast\mathcal{G}
\wedge \mathcal{G}^{T}+\ast \mathcal{G}\wedge
\mathcal{G})+c_{2}e^{\alpha_{1}\sigma} \ast F\wedge\mathcal{M} F.
\end{equation}
Furthermore by using \eqref{294.6} we have
\begin{equation}\label{304}
\begin{aligned}
\mathcal{L}(\phi^{i},\chi^{m})=&-\frac{1}{2}\sum\limits_{i=1}^{r}\ast d\phi ^{i}\wedge d%
\phi ^{i}-\frac{1}{2}\sum\limits_{\beta\in\Delta_{nc}^{+}}e^{\beta
_{i}\phi ^{i}}\ast U^{\beta }\wedge U^{\beta
}\\
\\
&+c_{2}e^{\alpha_{1}\sigma} \ast F\wedge\mathcal{M} F.
\end{aligned}
\end{equation}
Following the outline of \cite{ker1,ker2} which treat $U^{\alpha}$
as independent fields and then use the Lagrange multiplier
formalism the field equations of $\phi^{i}$ and $\chi^{m}$ arising
from the variation of \eqref{304} have been derived in
\cite{nej3}. They can be given as
\begin{equation}\label{305}
\begin{aligned}
d(e^{\frac{1}{2}\gamma _{i}\phi ^{i}}\ast U^{\gamma
})&=-\frac{1}{2}\gamma _{j}e^{\frac{1}{2}\gamma _{i}\phi
^{i}}d\phi ^{j}\wedge \ast U^{\gamma }\\
\\
&\quad +\sum\limits_{\tau -\beta =-\gamma }e^{\frac{1}{2} \tau
_{i}\phi ^{i}}e^{\frac{1}{2}\beta _{i}\phi ^{i}}N_{\tau ,-\beta
}U^{\tau }\wedge \ast U^{\beta
},\\
\\
d(\ast d\phi ^{i})&=\frac{1}{2}\sum\limits_{\beta\in\Delta_{nc}^{+}}^{}\beta _{i}%
e^{\frac{1}{2}\beta _{j}\phi ^{j}}U^{\beta }\wedge e^{%
\frac{1}{2}\beta _{j}\phi ^{j}}\ast U^{\beta
}\\
\\
&\quad
+(-1)^{D}c_{2}e^{\alpha_{1}\sigma}((H_{i})_{N}^{A}\nu_{M}^{N}\nu_{J}^{A})\ast
F^{M}\wedge F^{J},
\end{aligned}
\end{equation}
$\tau,\beta,\gamma\in\Delta_{nc}^{+}$ and $i=1,...,$
dim($\mathbf{h}_{k})\equiv r$. The matrices $\{(H_{i})_{N}^{A}\}$
are the ones corresponding to the generators $\{H_{i}\}$ in the
($20-2D+N$)-dimensional representation chosen. By also referring
to \cite{nej1} the field equations in \eqref{305} can more
compactly be written as
\begin{equation}\label{306}
\begin{aligned}
d(e^{\gamma _{i}\phi ^{i}}\ast U^{\gamma })&=\sum\limits_{\alpha
-\beta=-\gamma }N_{\alpha ,-\beta }U^{\alpha }\wedge e^{\beta
_{i}\phi ^{i}}\ast U^{\beta },\\
\\
d(\ast d\phi ^{i})&=\frac{1}{2}\sum\limits_{\beta\in\Delta_{nc}^{+}}^{}\beta _{i}%
e^{\frac{1}{2}\beta _{j}\phi ^{j}}U^{\beta }\wedge e^{%
\frac{1}{2}\beta _{j}\phi ^{j}}\ast U^{\beta
}\\
\\
&\quad +(-1)^{D}c_{2}e^{\alpha_{1}\sigma}\ast
F\wedge\nu^{T}H_{i}\nu F.
\end{aligned}
\end{equation}
It is worth stating that the last term in the second equation
above arises from the variation of the matter lagrangian
\eqref{211.5} with respect to the dilaton $\phi^{i}$. A similar
term is missing for the axions $\chi^{m}$ owing to the Lagrange
multiplier method used in the derivation of the corresponding
field equations \cite{ker1,nej1}. Therefore we have
\begin{equation}\label{307}
(-1)^{D}c_{2}e^{\alpha_{1}\sigma}\ast F\wedge\nu^{T}H_{i}\nu
F=(-1)^{D}\frac{\partial\mathcal{L}_{m}}{\partial\phi^{i}}.
\end{equation}
From \eqref{211.5}
\begin{equation}\label{308}
\begin{aligned}
\frac{\partial\mathcal{L}_{m}}{\partial\phi^{l}}&=\frac{\partial
(c_{2}e^{\alpha_{1}\sigma}\mathcal{M}_{IJ} \ast F^{I}\wedge
F^{J})}{\partial\phi^{l}}\\
\\
&=\frac{\partial (c_{2}e^{\alpha_{1}\sigma}\nu_{I}^{A}\nu_{J}^{A}
\ast F^{I}\wedge F^{J})}{\partial\phi^{l}}\\
\\
&=c_{2}e^{\alpha_{1}\sigma}(\partial_{l}\nu_{I}^{A}\nu_{J}^{A}+\nu_{I}^{A}\partial_{l}\nu_{J}^{A})\ast
F^{I}\wedge F^{J}.
\end{aligned}
\end{equation}
Since in the parametrization \eqref{23} the dilatons $\phi^{i}$
are coupled to the Cartan generators $H_{i}$ which commute among
themselves within the Cartan subalgebra structure of
$o(10-D,10-D+N)$ one can easily show that
\begin{equation}\label{309}
\frac{\partial\nu}{\partial\phi^{l}}\equiv\partial_{l}\nu=\frac{1}{2}H_{l}\nu.
\end{equation}
Thus we have
\begin{equation}\label{309}
\partial_{l}\nu_{I}^{A}=(\partial_{l}\nu)_{I}^{A}=(\frac{1}{2}H_{l}\nu)_{I}^{A}=\frac{1}{2}(H_{l})_{N}^{A}\nu_{I}^{N}.
\end{equation}
Inserting this result back in \eqref{308} we obtain
\begin{equation}\label{310}
\frac{\partial\mathcal{L}_{m}}{\partial\phi^{l}}=
c_{2}e^{\alpha_{1}\sigma}(\frac{1}{2}(H_{l})_{N}^{A}\nu_{I}^{N}\nu_{J}^{A}+\frac{1}{2}\nu_{I}^{A}(H_{l})_{N}^{A}\nu_{J}^{N})\ast
F^{I}\wedge F^{J}.
\end{equation}
Now since both $F^{I}$ and $F^{J}$ are two-forms by using the
identity
\begin{equation}\label{311}
(H_{l})_{N}^{A}\nu_{I}^{N}\nu_{J}^{A}\ast F^{I}\wedge
F^{J}=(H_{l})_{N}^{A}\nu_{I}^{N}\nu_{J}^{A}\ast F^{J}\wedge F^{I},
\end{equation}
and by arranging the indices above we conclude that
\begin{equation}\label{310}
\begin{aligned}
\frac{\partial\mathcal{L}_{m}}{\partial\phi^{l}}&=
c_{2}e^{\alpha_{1}\sigma}\nu_{I}^{A}(H_{l})_{N}^{A}\nu_{J}^{N}\ast
F^{I}\wedge F^{J}\\
\\
&=c_{2}e^{\alpha_{1}\sigma}\ast F\wedge\nu^{T}H_{l}\nu F,
\end{aligned}
\end{equation}
which justifies its existence in \eqref{306}. In this section we
have derived the set of equations \eqref{211.6}, \eqref{211.8},
\eqref{284} and \eqref{306} which are the field equations of the
$O(p,q)/O(p)\times O(q)$ symmetric space sigma model coupled
generally to a dilaton, a certain number of abelian gauge fields
and a two-form which is coupled through its Chern-Simons field
strength. We have shown that the equations \eqref{284} can be
expressed more explicitly in terms of the coset scalars through
the calculation of the Cartan-Maurer form. Thus we have also
derived a simplified set of equations namely the equations
\eqref{293} as alternatives of \eqref{284}. The equations
\eqref{305} are also written more compactly in \eqref{306}. Before
we link our results to the $D$-dimensional toroidally compactified
heterotic string we will give three examples on which the results
we have obtained are applicable.
\subsection{The D=7 Case}
The bosonic lagrangian of the $\mathcal{N}=2$, $D=7$
Maxwell-Einstein supergravity is constructed in \cite{d=7} as
\begin{equation}\label{yenii1}
\begin{aligned}
\mathcal{L}&=\frac{1}{2}R\ast1-\frac{5}{8} \ast d\sigma\wedge
d\sigma-\frac{1}{2}e^{2\sigma}\ast G
\wedge G\\
\\
&\quad +\frac{1}{4}tr( \ast d\mathcal{M}^{-1}\wedge
d\mathcal{M})-\frac{1}{2}e^{\sigma} F\wedge\mathcal{M} \ast F,
\end{aligned}
\end{equation}
where the coupling between the field strengths $F^{I}=dA^{I}$ for
$I=1,...,N+3$ and the scalars which parameterize the coset
$SO(N,3)/SO(N)\times SO(3)$ can be explicitly written as
\begin{equation}\label{yenii2}
-\frac{1}{2}e^{\sigma} F\wedge\mathcal{M}\ast
F=-\frac{1}{2}e^{\sigma}\mathcal{M}_{IJ} F^{I}\wedge \ast F^{J}.
\end{equation}
In writing \eqref{yenii1} we have assumed the ($N+3$)-dimensional
matrix representation of $so(N,3)$. The Chern-Simons form $G$ is
defined as \cite{d=7}
\begin{equation}\label{yenii3}
 G=dB-\frac{1}{\sqrt{2}}\:\eta_{IJ}\:A^{I}\wedge F^{J},
\end{equation}
where $\eta=$ diag$(-,-,-,+,+,...,+)$ is the invariant tensor of
$SO(N,3)$. If we compare \eqref{yenii1} with \eqref{210} we see
that
\begin{subequations}
\begin{gather}\label{yenii4}
c_{1}=-\frac{5}{8}\quad ,\quad c_{2}=c_{3}=-\frac{1}{2},\notag\\
 \notag\\
 \alpha_{1}=1\quad ,\quad\alpha_{2}=2\tag{\ref{yenii4}}.
\end{gather}
\end{subequations}
Also comparing \eqref{yenii3} with \eqref{212} yields
\begin{equation}\label{yenii5}
c_{4}=-\frac{1}{\sqrt{2}}.
\end{equation}
Thus with the above identification of the coefficients
\eqref{211.6}, \eqref{211.8}, \eqref{284} and \eqref{306} give us
the bosonic field equations of the $\mathcal{N}=2$, $D=7$
Maxwell-Einstein supergravity in a general solvable Lie algebra
parametrization of the coset $SO(N,3)/SO(N)\times SO(3)$. We
should state that although this coset is different than the one we
have assumed in \eqref{yeni5} the formal, abstract construction of
the lagrangian in the solvable Lie algebra gauge we have performed
for \eqref{yeni5} is also valid for $SO(N,3)/SO(N)\times SO(3)$.
\subsection{The D=8 Case}
The lagrangian of the bosonic sector of the $\mathcal{N}=1$, $D=8$
Maxwell-Einstein supergravity is given as \cite{d=8}
\begin{equation}\label{yenii6}
\begin{aligned}
\mathcal{L}&=\frac{1}{4}R\ast1+\frac{3}{8} \ast d\sigma\wedge
d\sigma-\frac{1}{2}e^{2\sigma} \ast G
\wedge G\\
\\
&\quad+\frac{1}{4}tr( \ast d\mathcal{M}^{-1}\wedge
d\mathcal{M})-\frac{1}{2}e^{\sigma} F\wedge\mathcal{M} \ast F.
\end{aligned}
\end{equation}
The Chern-Simons three-form $G$ is
\begin{equation}\label{yenii7}
 G=dB+\eta_{IJ}F^{I}\wedge A^{J},
\end{equation}
with $I,J=1,...,N+2$. Apart from the single dilaton the rest of
the scalars parameterize the coset manifold $SO(N,2)/SO(N)\times
SO(2)$ and the $SO(N,2)$ invariant tensor $\eta$ is $\eta=$
diag$(-,-,+,+,...,+)$. We again assume that we choose an
($N+2$)-dimensional matrix representation of $so(N,2)$. Again if
we compare \eqref{yenii6} with \eqref{210} in this case we have
\begin{subequations}
\begin{gather}\label{yenii8}
c_{1}=\frac{3}{8}\quad ,\quad c_{2}=c_{3}=-\frac{1}{2},\notag\\
 \notag\\
 \alpha_{1}=1\quad ,\quad\alpha_{2}=2\tag{\ref{yenii8}}.
\end{gather}
\end{subequations}
Also from \eqref{yenii7} and \eqref{212} we have
\begin{equation}\label{yenii5}
c_{4}=1.
\end{equation}
If the coefficients are chosen as above then \eqref{211.6},
\eqref{211.8}, \eqref{284} and \eqref{306} correspond to the
bosonic field equations of the $\mathcal{N}=1$, $D=8$
Maxwell-Einstein supergravity in a general solvable Lie algebra
parametrization of the coset $SO(N,2)/SO(N)\times SO(2)$.
\subsection{The D=9 Case}
As a final example we will consider the $\mathcal{N}=1$, $D=9$
Maxwell-Einstein supergravity \cite{d=9}. The bosonic lagrangian
of the $\mathcal{N}=1$, $D=9$ Maxwell-Einstein supergravity can be
given as \cite{d=9}
\begin{equation}\label{yenii9}
\begin{aligned}
\mathcal{L}&=-\frac{1}{4}R\ast1+\frac{7}{4} \ast d\sigma\wedge
d\sigma+\frac{1}{2}e^{-4\sigma}\ast G
\wedge G\\
\\
&\quad +\frac{1}{4}tr( \ast d\mathcal{M}^{-1}\wedge
d\mathcal{M})-\frac{1}{2}e^{-2\sigma} F\wedge\mathcal{M} \ast F.
\end{aligned}
\end{equation}
In this case the scalars of the coupling abelian vector multiplets
parameterize the scalar coset $SO(N,1)/SO(N)$. The Chern-Simons
three-form is taken as
\begin{equation}\label{yenii10}
 G=dB+\eta_{IJ}A^{I}\wedge F^{J},
\end{equation}
with $I,J=1,...,N+1$. We assume an ($N+1$)-dimensional matrix
representation of $so(N,1)$. Similar to the $D=7$ and $D=8$ cases
we have $\eta=$ diag$(-,+,+,...,+)$ which is the invariant tensor
of $SO(N,1)$. The comparison of \eqref{yenii9} and \eqref{yenii10}
with \eqref{210} and \eqref{212} respectively specifies the
coefficients as
\begin{subequations}
\begin{gather}\label{yenii11}
c_{1}=\frac{7}{4}\quad ,\quad c_{2}=-\frac{1}{2}\quad ,\quad c_{3}=\frac{1}{2},\notag\\
 \notag\\
 \alpha_{1}=-2\quad ,\quad\alpha_{2}=-4\quad ,\quad c_{4}=1\tag{\ref{yenii11}}.
\end{gather}
\end{subequations}
Therefore again with the above identification of the coefficients
\eqref{211.6}, \eqref{211.8}, \eqref{284} and \eqref{306} become
the bosonic field equations of the $\mathcal{N}=1$, $D=9$
Maxwell-Einstein supergravity for a general solvable Lie algebra
parametrization of the coset $SO(N,1)/SO(N)$. We should remark
that the discussion we have made for the replacement of
\eqref{yeni5} with the coset $SO(N,3)/SO(N)\times SO(3)$ of the
$D=7$ case is also valid for the $D=8$ coset $SO(N,2)/SO(N)\times
SO(2)$ and the $D=9$ coset $SO(N,1)/SO(N)$.
\section{Field Equations of the D-dimensional Heterotic String}
In section two we have given the bosonic lagrangian of the
toroidally compactified $E_{8}\times E_{8}$ low energy effective
heterotic string. The Kaluza-Klein reduction produces the scalar
manifolds which are in the form of \eqref{ch2319}. The scalar
lagrangian is written as \eqref{29.5} by using the explicit
representation \eqref{ch2338} of the generators of the solvable
Lie algebra of $o^{\prime}(10-D+N,10-D)$ in the coset
parametrization \eqref{ch2332}. As we have already mentioned
before the importance of this formulation is that it is based on
the Kaluza-Klein descendant fields. In section three however we
have taken our scalar coset manifold explicitly as
$O(10-D,10-D+N)/O(10-D)\times O(10-D+N)$ whose definition
originates from the indefinite signature metric \eqref{24.5}. We
have to remark two points now. Firstly in the definition of the
generalized orthogonal groups $O(p,q)$ the indefinite signature
metric can be replaced by any symmetric matrix which has $p$
positive and $q$ negative eigenvalues. The resulting group of
matrices may differ from each other from their set content point
of view however their group structures are all isomorphic to each
other \cite{knapp}. Also since $Spin(p,q)$ is a double cover of
$O(p,q)$ and since it is isomorphic to $Spin(q,p)$ \cite{cli} we
have the isomorphisms
\begin{equation}\label{41}
\begin{aligned}
&O^{\prime}(10-D+N,10-D)\simeq O^{\prime}(10-D,10-D+N)\\
\\
&\simeq O(10-D+N,10-D)\simeq O(10-D,10-D+N).
\end{aligned}
\end{equation}
Thus we observe that from the algebraic point of view the choice
of the global symmetry group representation among \eqref{41} which
defines the scalar coset manifold is unimportant. However the
choice of $O^{\prime}(10-D+N,10-D)$ is distinct since it arises
naturally during the Kaluza-Klein reduction in \cite{heterotic}.
On the other hand the choice of the global symmetry group
representation effects the structure of the field equations via
the indefinite metric which explicitly appears in the equations.
The second point we should emphasize on is that the properties of
$\eta$ which we have made use of to derive the bosonic field
equations of the scalar coset with couplings in the previous
section are all shared by $\Omega$. First of all as it can be
deduced directly from \eqref{ch2343}
\begin{equation}\label{42}
\Omega^{2}=\mathbf{1}.
\end{equation}
Also since $\Omega$ is a symmetric matrix and the coset
representatives \eqref{ch2332} of
$O^{\prime}(10-D+N,10-D)/O(10-D+N)\times O(10-D)$ satisfy
\eqref{29.1} the identities \eqref{287}-\eqref{289.5} are shared
by $\Omega$ and the coset representatives \eqref{ch2332}.
Therefore we may conclude that we may replace $\eta$ by $\Omega$
in the formulation performed in the last section, thus our results
can be extended to the Kaluza-Klein scalar coset representation of
section two. In other words all the formulas of section three are
legitimately correct when one replaces $\eta$ by $\Omega$ taking
the scalar coset manifold as
$O^{\prime}(10-D+N,10-D)/O(10-D+N)\times O(10-D)$. One final
remark we should do before writing down the field equations of the
heterotic string is about the coset parametrization. Comparing the
coset parametrizations \eqref{ch2332} and \eqref{23} which are
both based on the solvable Lie algebra of the global symmetry
group one sees that one needs a transformation to relate the
fields $\{\varphi^{i},A^{i}_{(0)j},A_{(0)ij},B^{I}_{(0)i}\}$ to
the ones $\{\phi^{i},\chi^{m}\}$. Thus one should consider the
equality of the two coset parametrizations
\begin{equation}\label{43}
e^{\frac{1}{2}\varphi^{i}H_{i}}e^{A^{i}_{(0)j}E_{i}^{j}}e^{\frac{1}{2}A_{(0)ij}
V^{ij}}e^{B^{I}_{(0)i}U_{I}^{i}}=e^{\frac{1}{2}\phi
^{i}H_{i}}e^{\chi ^{m}E_{m}},
\end{equation}
bearing in mind that while the solvable Lie algebra generators on
the left hand side are fixed by the representation chosen and by
their embedding in $o^{\prime}(10-D+N,10-D)$ which is discussed in
\cite{heterotic}, the solvable Lie algebra generators on the right
hand side are completely arbitrary. The transformation can be
calculated in two ways; either after choosing the basis on both
sides of \eqref{43} the same, one considers local scalar fields
whose ranges in the exponential generate solvable Lie algebra
elements in a sufficiently small neighborhood of the identity
element so that \eqref{43} is valid in any representation chosen
for the algebra, thus giving the transformations explicitly as a
matrix equation $\footnote{A similar discussion leads to the
differential form of such scalar field transformations in
\cite{nej2}.}$ or after determining another solvable Lie algebra
$\mathbf{s}_{0}$ of $o^{\prime}(10-D+N,10-D)$ through a selection
of an Iwasawa decomposition on the right hand side of \eqref{43},
one may identify the matrix representatives of any basis of this
solvable Lie algebra $\{H_{i},E_{m}\}$ in the representation
generated by \eqref{ch2338} then one can find the field
transformations from \eqref{43} explicitly again. Despite the
locality of the first method by following the second method one
may find global field transformations which depend on the
particular representation of section two. Therefore keeping in
mind that one can always find such transformations between the
scalar field definitions of the two different coset
parametrizations as discussed above, in writing the field
equations of the $D$-dimensional compactified heterotic string we
will assume a coset parametrization in the form of \eqref{23} for
the scalar coset manifold $O^{\prime}(10-D+N,10-D)/O(10-D+N)\times
O(10-D)$ which is the one used in the previous section. Comparing
the lagrangians \eqref{ch2339} and \eqref{210} we firstly observe
that
\begin{equation}\label{43.5}
\eta\rightarrow\Omega.
\end{equation}
Then
\begin{subequations}
\begin{gather}\label{44}
c_{1}=c_{2}=c_{3}=-\frac{1}{2},\notag\\
 \notag\\
 \alpha_{1}=-\sqrt{\frac{2}{(D-2)}},\quad\alpha_{2}=-\sqrt{\frac{8}{(D-2)}}\tag{\ref{44}}.
\end{gather}
\end{subequations}
Also
\begin{subequations}
\begin{gather}\label{45}
\sigma\equiv\phi,\quad G\equiv F_{(3)},\quad C_{(1)}^{I}\equiv
A^{I},\notag\\
 \notag\\
 B\equiv A_{(2)},\quad F^{I}\equiv H_{(2)}^{I}.\tag{\ref{45}}
\end{gather}
\end{subequations}
Furthermore
\begin{equation}\label{46}
\begin{aligned}
F_{(3)}&=dA_{(2)}+\frac{1}{2}C_{(1)}^{T}\:\Omega\: dC_{(1)}\\
\\
&=dA_{(2)}+\frac{1}{2}\:\Omega_{IJ}\: C_{(1)}^{I}\wedge
dC_{(1)}^{J},
\end{aligned}
\end{equation}
which is in the form of \eqref{212} with the identification of the
coupling constant as
\begin{equation}\label{46.1}
c_{4}=\frac{1}{2}.
\end{equation}
We may conclude that in the light of these identifications the
formalism which appears in section two is in parallel with the
general one constructed in section three differing only in the
scalar coset representation. Finally upon the discussion we have
made about the coset parametrizations, if we consider the field
redefinitions and the substitutions mentioned above in the
formalism of the previous section a similar line of derivation
would give us the bosonic field equations of the $D$-dimensional
toroidally compactified $E_{8}\times E_{8}$ low energy effective
heterotic string as
\begin{equation}
\begin{aligned}
(-1)^{D}d(\ast
d\phi)&=\frac{1}{2}\sqrt{8/(D-2)}\:e^{-\sqrt{\frac{8}{(D-2)}}\phi}
\ast F_{(3)}\wedge
F_{(3)}\\
\\
&\quad
+\frac{1}{2}\sqrt{2/(D-2)}\:e^{-\sqrt{\frac{2}{(D-2)}}\phi}\mathcal{M}_{IJ}
\ast
H_{(2)}^{I}\wedge H_{(2)}^{J},\\
\\
d(e^{-\sqrt{\frac{8}{(D-2)}}\phi}\ast F_{(3)})&=0,\\
\\
d(e^{-\sqrt{\frac{2}{(D-2)}}\phi}\mathcal{M}\ast
H_{(2)})&=(-1)^{D}e^{-\sqrt{\frac{8}{(D-2)}}\phi}\Omega H_{(2)}\wedge\ast F_{(3)},\\
\\
d(\ast d\phi ^{i})&=\frac{1}{2}\sum\limits_{\beta\in\Delta_{nc}^{+}}^{}\beta _{i}%
e^{\frac{1}{2}\beta _{j}\phi ^{j}}U^{\beta }\wedge e^{%
\frac{1}{2}\beta _{j}\phi ^{j}}\ast U^{\beta
}\\
\\
&\quad -\frac{1}{2}(-1)^{D}e^{-\sqrt{\frac{2}{(D-2)}}\phi}\ast
H_{(2)}\wedge\nu^{T}H_{i}\nu H_{(2)},\notag
\end{aligned}
\end{equation}
\begin{equation}\label{47}
\begin{aligned}
 d(e^{\gamma _{i}\phi
^{i}}\ast U^{\gamma })&=\sum\limits_{\alpha -\beta=-\gamma
}N_{\alpha ,-\beta }U^{\alpha }\wedge e^{\beta _{i}\phi ^{i}}\ast
U^{\beta },
\end{aligned}
\end{equation}
where the coset representatives
\begin{equation}\label{48}
\nu=e^{\frac{1}{2}\phi ^{i}H_{i}}e^{\chi ^{m}E_{m}},
\end{equation}
satisfy the defining relation of $O^{\prime}(10-D+N,10-D)$ namely
\begin{equation}\label{49}
\nu^{T}\Omega \nu=\Omega,
\end{equation}
and they parameterize the coset
\begin{equation}\label{410}
O^{\prime}(10-D+N,10-D)/O(10-D+N)\times O(10-D),
\end{equation}
as we have discussed in detail above. Before concluding we will
discuss one more point. One may investigate the relation between
the two formulations considered in this section also in section
two and the one introduced in section three which are constructed
over different scalar coset representations although they are two
equivalent formulations of the same theory. This is a matter of
interest since when we take a look at \eqref{46} and \eqref{47} we
see that as $\Omega$ is a non-diagonal matrix there is a degree of
mixing of the abelian gauge fields and their field strengths in
the Chern-Simons three-form \eqref{46}. This mixing brings a
certain complexity of coupling in the field equations \eqref{47}.
However in the formulation of section three this mixing is avoided
since $\eta$ is a diagonal matrix. Thus the field equations
obtained by using the scalar coset representation based on $\eta$
involve less coupling which may be a simplification and an
advantage in seeking solutions. One may find solutions in the
$\eta$-formulation and then pass to the Kaluza-Klein field content
by using the field transformations upon the assumption that the
lagrangians of these two distinct formulations are equal. This
approach is different than the one used above which derives the
field equations of the $D$-dimensional heterotic string solely in
the $\Omega$-formulation. More specifically one would consider the
equality of the two lagrangians which are based on the scalar
coset manifolds of $O^{\prime}(10-D+N,10-D)/O(10-D+N)\times
O(10-D)$ and $O(10-D,10-D+N)/O(10-D)\times O(10-D+N)$ respectively
\footnote{Note that although we have used the first coset
structure for the $D$-dimensional heterotic string in section two
and four and the second one in our general formalism in section
three we have made use of the results of section three for the
$D$-dimensional heterotic string by simply substituting $\Omega$
instead of $\eta$ owing to the similar properties of these two
metrics which take part in the derivation of the field equations.
However we have not mentioned relating these two formulations
before.}. In this case one has to calculate the field
transformations between the constructions based on two different
scalar coset representations. We may calculate the transformations
of the field contents by equating the two lagrangians
\eqref{ch2339} and \eqref{210} which are based on different scalar
coset matrix structures. If we do so we see that all the previous
coupling constant and field equivalences mentioned in this section
are valid except the ones for $A^{I}$ and the coset scalars since
in this case we have
\begin{equation}\label{411}
\frac{1}{2}\:\Omega_{IJ}\: C_{(1)}^{I}\wedge
dC_{(1)}^{J}=c_{4}\eta_{IJ}A^{I}\wedge F^{J},
\end{equation}
and
\begin{equation}\label{412}
-\frac{1}{2}e^{-\sqrt{2/(D-2)}\phi}\ast
H_{(2)}^{T}\wedge{\mathcal{M}}H_{(2)}=c_{2}e^{\alpha_{1}\sigma}
\ast F\wedge\mathcal{M} F.
\end{equation}
On the left hand side of the above equation the coset
parametrization is \eqref{ch2332} and on the right hand side it is
\eqref{23}. If one assumes a transformation between the gauge
fields $C_{(1)}^{I}$ and $A^{I}$ of the two formulations in the
form
\begin{equation}\label{413}
C_{(1)}=TA,
\end{equation}
where $T$ is a ($20-2D+N$)$\times$($20-2D+N$) matrix then by
choosing $c_{4}=1/2$ again one finds that from \eqref{411} $T$
must satisfy
\begin{equation}\label{414}
T^{T}\Omega T=\eta.
\end{equation}
Inserting \eqref{413} in \eqref{412} one also finds that
\begin{equation}\label{415}
\nu^{\prime}=\nu T.
\end{equation}
Since $\Omega$ is a symmetric matrix and since any symmetric
matrix can be diagonalized in the form \eqref{414} with $T\in
O(20-2D+N)$ \cite{kunze,schutz,cornwell} one can find the
transformation matrix $T$ to relate the two different set of gauge
fields. After determining the solvable Lie algebra
$\mathbf{s}_{0}$ of $O(10-D,10-D+N)$ through a selection of an
Iwasawa decomposition \eqref{21} one may choose a
($20-2D+N)$-dimensional matrix representation to express the basis
$\{H_{i},E_{m}\}$ and then from \eqref{415} one may also find the
transformation rules of the coset scalars between the two scalar
coset formulations. In this way one may solve the field equations
in the $\eta$-formulation which are less coupled and then pass to
the field content obtained by the direct Kaluza-Klein reduction.

Although this method may look tempting we should also point out
that the transformations obtained in this way can not cover the
entire solution space thus they may be highly restrictive. The
reason for this lies in the process of equating the two
lagrangians \eqref{ch2339} and \eqref{210} and obtaining
\eqref{415}. We have stated before that although the groups are
isomorphic the matrix contents of the sets
$O^{\prime}(10-D+N,10-D)$ and $O(10-D,10-D+N)$ may be quite
different for example
\begin{subequations}\label{416}
\begin{gather}
\Omega\in O^{\prime}(10-D+N,10-D),\quad \Omega\notin
O(10-D,10-D+N),\notag\\
\notag\\
\eta\in O(10-D,10-D+N),\quad \eta\notin
O^{\prime}(10-D+N,10-D).\notag\\\tag{\ref{416}}
\end{gather}
\end{subequations}
Thus the transformations \eqref{415} can exist only for the
localized coset scalars which would lead to the coset
representatives that lie in $O^{\prime}(10-D+N,10-D)$ when the
representatives in $O(10-D,10-D+N)$ are translated by $T$ from the
right. On the other hand due to the generalized formulation of
section three one has the degree of freedom of defining
$\{\phi^{i},\chi^{m}\}$ from a rich choice of alternatives of the
Iwasawa decomposition, the solvable Lie algebra, the basis
$\{H_{i},E_{m}\}$ and the matrix representations of the Lie
algebra $o(10-D,10-D+N)$. Finally we may state that to justify the
transformations \eqref{415} one should focus on matching the coset
representative images of the solvable Lie algebra parametrizations
of the two distinct scalar coset structures of section two and
section three via the right-translation by $T$.
\section{Conclusion}
A review of the Kaluza-Klein dimensional reduction
\cite{heterotic} of the bosonic sector of the ten dimensional
$\mathcal{N=}1$ simple supergravity that is coupled to $N$ abelian
gauge multiplets \cite{d=10,tani15} on the tori $T^{10-D}$ which
describes the massless sector of the $E_{8}\times E_{8}$ heterotic
string theory \cite{kiritsis} is given in section two. Then we
have derived the field equations of the general symmetric space
sigma model with dilaton, abelian matter and Chern-Simons
couplings in section three. In section four we have used the
results of section three to express the bosonic field equations of
the $D$-dimensional heterotic string whose bosonic lagrangian is
in parallel with the general lagrangian we have considered in
section three. Besides as we have discussed by the end of section
three our general formulation corresponds to the one that occurs
when the Maxwell-Einstein supergravities are constructed by the
use of the Noether's method \cite{d=7,d=8,d=9}. We have also
mentioned about the possible transformations between the two
different scalar coset parametrizations of section two and section
three which are both in the solvable Lie algebra gauge. Finally in
section four we have discussed an alternative way of deriving the
field equations of the compactified heterotic string whose field
content directly arises from the $D=10$ heterotic string via the
Kaluza-Klein ansatz \cite{heterotic}. By this method one can
express the field equations of the $D$-dimensional heterotic
string exactly as in section three whose scalar coset matrix
representations are different from the ones used in section two.
One can then search for the field transformations which originate
from the equality of the two distinct lagrangians of section two
and section three. The limitations associated with this method are
also pointed out in the last section.

In \cite{nej3} the dynamics of the symmetric space sigma model
with only abelian gauge field couplings is studied under the
solvable Lie algebra gauge. Therefore apart from deriving the
bosonic matter field equations of the $D$-dimensional $E_{8}\times
E_{8}$ heterotic string this work generalizes the formulation of
\cite{nej3} to include further  dilaton and Chern-Simons
couplings. Our formulation in section three is based on a generic
global symmetry group $O(p,q)$ and we derive the field equations
for a general solvable Lie algebra parametrization of the scalar
coset without specifying neither the basis used nor the
representation chosen as in \cite{nej3}. From this point of view
the field equations we have obtained are extensive in their
coverage and they are purely in algebraic terms. On the other hand
the major achievement of this work is the explicit derivation of
the bosonic matter field equations of the $D$-dimensional
toroidally compactified $E_{8}\times E_{8}$ heterotic string for a
generic and unspecified solvable Lie algebra parametrization of
the scalar coset. Although the bosonic lagrangian constructed in
\cite{heterotic} under the solvable Lie algebra gauge assumes a
certain representation of the coset generators we have obtained
the bosonic matter field equations of the $D$-dimensional
heterotic string in a representation free formalism. Therefore for
the field equations obtained in the last section we have a degree
of freedom of choosing an appropriate representation of the
solvable Lie algebra. This would provide a useful machinery for
seeking solutions of the theory. As a matter of fact one can
inspect various representations to simplify the field equations
bearing in mind that choosing a basis for the solvable Lie algebra
also corresponds to defining the coset scalar fields. Appropriate
representations would decrease the complexity of the field
equations. When one finds a suitable representation step by step
one can construct the explicit equations of motion which would
involve less couplings. If the field equations in section three
and section four are inspected carefully one can see that the most
essential coupling is between the coset scalars and the gauge
fields beside the non-linearity of the structure of the coset
scalars themselves. Thus focusing on these couplings we have also
derived various versions of the corresponding field equations in
section three which may provide a set of tools when certain
solutions of these equations are studied.

In section four we have discussed that the formulation of section
three which is based on the invariant metric $\eta$ is
advantageous over the formulation descending from the Kaluza-Klein
compactification so that $\eta$ is a diagonal matrix and it
prevents cross couplings of the gauge fields in the field
equations. Since our aim in this work is to formulate the field
equations rather than studying the solutions we have not given
specific examples for which this simplicity may be exploited.
However we may state that the $\eta$-formulation brings simplicity
in deriving the first-order field equations of the
Maxwell-Einstein supergravities and in constructing the non-linear
realizations of these theories \cite{nej5,nej4,nej6}. For the
$\eta$-formulation from \cite{nej5,nej4,nej6} we observe that the
non-mixing of the gauge fields in the field equations is reflected
in the coset algebra of the non-linear sigma model construction of
the theory such that the operators which correspond to different
gauge fields do not mix under the algebra product (they
anti-commute with each other).

Although we have mentioned about the relation between the two
different solvable Lie algebra parametrizations of the scalar
coset in the last section we have not shown any attempt to
construct the transformations explicitly. As we have discussed one
can search for the local or the global scalar field
transformations which would make use of the group theoretical
structure of the global symmetry group. One can also inspect the
field transformations mentioned in the last part of section four
which relate the field contents of the two equivalent formulations
of the theory based on two distinct scalar coset representations.
As we have stated before this would remove some complexity of the
field equations. Therefore one can workout the solutions in the
formulation of section three and then transform the fields to the
original Kaluza-Klein field content of section two.

The formalism of section three can be extended for theories which
posses different global symmetry groups than $O(p,q)$ but which
still can pertain the solvable Lie algebra gauge. In this case the
scalar coset $G/K$ must be such that again the global symmetry
group $G$ must be a real form of a non-compact semi-simple Lie
group and $K$ must be its maximal compact subgroup to be able to
apply the solvable Lie algebra gauge to parameterize the scalar
coset manifold \cite{hel}. However when defining the internal
metric one must use a generalized transpose which is induced by
the Cartan involution of the Lie algebra $g$
\cite{hel,nej3,ker1,ker2,nej1,nej2} instead of the ordinary matrix
transpose. Since the supergravity theories give way to first-order
formulations \cite{julia1,julia2,julia3} the first-order
formulation of the field equations can be performed for both the
general formalism of section three and the $D$-dimensional
compactified heterotic string. Finally we should state that
although owing to the non-linear structure of the coset scalars
also the complexity of the coupling between the scalars and the
gauge fields we have focussed on the bosonic matter sector in this
work one can extend the solvable Lie algebra gauge formalism
developed here to derive the field equations in the presence of
the graviton and the fermionic sectors too.

\end{document}